\documentclass[aps,showpacs,manuscript,12pt]{revtex4}
\usepackage{amssymb}
\usepackage{amsmath}
\usepackage{graphicx}

\begin{document}

\title{\textbf{Plasma-sheath effects on the Debye screening problem}}
\author{D. Sarmah$^{1,2,3}$, M. Tessarotto$^{1,3}$ and M. Salimullah$^{4}$}
\affiliation{$^{1}$Department of Mathematics and Informatics,\\
University of Trieste, Trieste, Italy \\
$^{2}$ International Center for Theoretical Physics, ICPT/TRIL\\
Program, Trieste, Italy \\
$^{3}$ Consortium for Magnetofluid Dynamics, Trieste, Italy\\
$^{4}$ Department of Physics, Jahangirnagar University, Savar, Dhaka,\\
Bangladesh}

\begin{abstract}
The classical Debye-H\"{u}ckel screening effect of the electrostatic field
generated by isolated charged particles immersed in a plasma is reviewed.
The validity of the underlying mathematical model, and particularly of the
weak-field approximation, are analyzed. It is shown that the presence of the
plasma sheath around test particles and the resulting effect of charge
screening are essential for the description of plasmas which are strongly
coupled.
\end{abstract}

\date{\today }
\maketitle

\section{ Introduction}

In this work we intend to analyze the consistency of the traditional
mathematical model for the so-called \emph{Debye screening problem} (DSP)
originally formulated by Debye and H\"{u}ckel (\emph{DH model }\cite{Debye
1923}). In particular, we intend to prove that the 3D Poisson equation
involved in the DH model does not admit physically acceptable solutions,
i.e., solutions which are provided by ordinary functions and are at least
continuous in the domain of existence, i.e., are so-called classical (or
strong) solutions. For this purpose a modified model is proposed which takes
into account the effect of local plasma sheath (i.e., the local domain near
test particles where the plasma must be considered discrete). Basic
consequences of the model are discussed, which concern the asymptotic
properties of the solutions determined both for weakly and strongly-coupled
plasmas. As an application the charge screening effect in strongly-coupled
plasmas is investigated and an explicit expression of the effective charge
for the asymptotic DH potential is determined.\newline
An ubiquitous property of plasmas, either quasi-neutral or non-neutral and
weakly or strongly-coupled, is the so-called Debye shielding of the
electrostatic potential. This is generated by a single charged particle
(test particle) of the plasma, located at position $\mathbf{r}(t)$ and
belonging to the N-body system (with $N\gg 1$) of charged particles forming
a plasma, or more generally by a finite collection of test particles (with
positions $\mathbf{r}_{i}(t),$ $i=1,$ $N_{t},$ possibly with $N_{t}\ll N$).
The test particle and the plasma are both assumed non-relativistic when
referred to a suitable Galileian\ reference frame, i.e., defined in such a
way that $\left\vert \mathbf{V}\right\vert /c\ll 1,\left\vert \overset{\cdot
}{\mathbf{r}}(t)\right\vert /c\ll 1$, $\mathbf{V}$ and $\overset{\cdot }{%
\mathbf{r}}(t)$ being respectively the fluid velocity of the plasma and the
test particle velocity $\overset{\cdot }{\mathbf{r}}(t),$ both to be assumed
generally non-vanishing. It is well-known that this phenomenon occurs, in
particular, if the plasma is assumed suitably close to kinetic equilibrium,
namely if each particle species ($s$) is described by a Maxwellian kinetic
distribution function ($f_{Ms}$), carrying a finite number density $N_{s},$%
temperature $T_{s}$ and flow velocity $\mathbf{V}$. Here, $T_{s},\mathbf{V}$
and $N_{s}$ (the latter only in the absence of test particles) must result
constant or at least, in a suitable sense slowly dependent with respect to
position ($\mathbf{r}$) and time ($t$). \ As a consequence, sufficiently far
from each test particle, in a sense to be defined, the Coulomb electrostatic
potential results effectively screened and decays exponentially on a
characteristic scale specified by the Debye length ($\lambda _{D};$see
below). Of course, this phenomenology applies only provided local
perturbations of the kinetic distribution function are suitably small. In
such a case, due to spherical symmetry of the test particle, the screened
potential must result manifestly function only of the distance $\rho
=\left\vert \mathbf{r-r}(t)\right\vert $ between the point of measurement $%
\mathbf{r}$ and the position of the test particle $\mathbf{r}(t).$ In
addition, if both the test particle and the plasma are non-relativistic, the
potential must be also independent of the velocity of the test particle $%
\overset{\cdot }{\mathbf{r}}(t).$ This is the subject of the so-called Debye
screening problem, which regards the determination of the electrostatic
potential generated by test particles in a quasi-neutral plasma.\ The
mathematical model of DSP in his original formulation \cite{Debye 1923}
(recalled in detail below, see Sec.2) concerns \emph{point-like test
particles} and is based both on the neglect of the \emph{local plasma sheath}
around each test particle (or the finite size of the test particles, for
example in the case of large-size dust particles) and on the so-called \emph{%
weak-field approximation}. The first assumption, concerns the neglect of the
screening effect produced by the presence of discrete interactions close to
the test particles (local plasma sheath). This set is generally
species-dependent and is defined as the spherical domain $R_{\left\{ \rho
_{os}\right\} }=\left\{ \rho :\rho _{p}\leq \rho <\rho _{os},\rho \equiv
\left\vert \mathbf{r-r}(t)\right\vert \in
\mathbb{R}
^{+}\right\} ,$ $\rho _{p}$ being the radius of a spherically-symmetric test
particle and $\rho _{os}$ the characteristic radius of the plasma sheath
produced bt the $s-$th particle species. The characteristic radius $\rho
_{os}$ can manifestly be identified with the mean interparticle distance in
the plasma for the species $s$, i.e., $\rho _{os}=\left( \frac{3}{4\pi N_{os}%
}\right) ^{1/3},$\ $N_{os}$ denoting the $s-$th species plasma number
density in the absence of test particles. This set is centered at the
position of the test particle, $\mathbf{r}(t),$ in which the species $s$ of
the plasma must be treated as a discrete system; instead, for any distance $%
\rho \geq \rho _{os}$ the same species is assumed to be described in terms
of a continuous kinetic distribution function. The second approximation,
which permits the linearization of the Poisson equation for the
electrostatic potential, is based on the requirement that, for arbitrary $%
\rho \equiv \left\vert \mathbf{r-r}(t)\right\vert \in
\mathbb{R}
^{+},$ the following asymptotic ordering holds
\begin{equation}
\widehat{\Phi }(\rho )\sim O(\varepsilon )\ll 1,  \label{eq-1-a}
\end{equation}
$\varepsilon $ being an infinitesimal and $\widehat{\Phi }=\left\vert e\Phi
\right\vert /T_{e}$ the normalized electrostatic potential and $T_{e}$ the
electron temperature measured in energy units. As a result, Debye and H\"{u}%
ckel obtained, for the electrostatic potential generated by a point-particle
carrying the electric charge $q,$ the asymptotic solution%
\begin{equation}
\Phi (\rho )\cong \Phi _{o}(\rho )\equiv \frac{\widehat{c}}{4\pi \left\vert
\mathbf{r-r}(t)\right\vert }e^{-\left\vert \mathbf{r-r}(t)\right\vert
/\lambda _{D}}  \label{Eq.1-bb}
\end{equation}%
which is the so-called \emph{DH potential}, also known as the Yukawa
potential. \ Here $\widehat{c}$ is the \emph{DH effective charge} to be
approximated
\begin{equation}
\widehat{c}\cong q,  \label{Eq.1-bbb}
\end{equation}%
and $\lambda _{D}$ is the Debye length
\begin{equation}
\frac{1}{\lambda _{D}^{2}}=\sum\limits_{j}\frac{1}{\lambda
_{Dj}^{2}}, \label{Eq.2-a}
\end{equation}%
where the summation is carried out over all plasma species $j.$ Furthermore,
$\lambda _{Dj}^{-1}=\sqrt{\frac{4\pi Z_{j}^{2}e^{2}N_{oj}}{T_{j}}}$ is the $%
j-th$ species Debye length and $N_{oj}$ the $j-th$ \ species plasma density
defined in the absence of test particles (i.e., letting $\widehat{\Phi }%
(\rho )\equiv 0)$.

Despite previous attempts to construct approximate or exact solutions to the
DH model \cite{Perez 1998,Chang Lin 2000,Martin 1994}, the related
mathematical model appears incomplete and can be shown to be physically
unacceptable, due to the neglect of the local plasma sheath. In fact, it is
obvious that sufficiently close to the point-particle the weak-field
approximation (\ref{eq-1-a}) is violated making the DH model invalid. In the
past \cite{Lampert 1980} it was pointed out that in such a case the test
particle does not produce any electric field, but only complete charge
neutralization by the plasma, thus producing a Debye length which
effectively vanishes. Other objections concerned the asserted indeterminacy
of the solution for $x=0$ due to its divergence in the same point \cite{Lo
Surdo 1994}. These issues were later addressed in a more general context
\cite{Garrett 1988}, including the 2D case where complete neutralization
cannot be achieved. To recover the correct physical picture the effect of
local plasma sheath must be included. Nevertheless, for suitably dense
plasmas or in the case of plasma species characterized by very high electric
charges (high-Z), such as dusty plasmas, the weak-field approximation may be
locally violated. This circumstance, when the effect of finite local plasma
sheath is included, occurs if the normalized electrostatic potential $%
\widehat{\Phi }(\rho )$ results of order unit or larger on the boundary of
the plasma sheath (produced by at least one of the $s$ plasma species),
namely for $\rho =\rho _{os}.$ Such a condition can be expressed in terms of
the Coulomb coupling parameter $\Gamma _{ks}$, \ i.e., by the requirement
that for a test particle of species $k$ there results at least for a species
$s$%
\begin{equation}
\Gamma _{ks}\equiv \frac{\beta _{k}}{x_{os}}=\frac{Z_{k}e^{2}}{T_{e}\rho
_{os}}\sim 1  \label{Eq.1}
\end{equation}%
(\emph{strong coupling condition}). Here the notation is standard. Thus, $%
\beta _{k}=\frac{Z_{k}e^{2}}{T_{e}\lambda _{D}},$ $\rho _{os}=\left( \frac{3%
}{4\pi N_{os}}\right) ^{1/3}$ and $x_{os}=\rho _{os}/\lambda _{D},$\ denote
respectively the dimensionless electric charge carried by a test particle of
species $k$ and the radius of the $s-th$ species plasma sheath surrounding
the test particle. We stress that, ordinarily, in plasmas the asymptotic
condition $x_{os}\ll 1$ has also to be imposed on each particle species $s$.
\ For example, in a space dusty plasma typical values for dust grain
charges, plasma temperature and plasma density are $Z_{d}\sim 10^{4}\div
10^{5},$ $T\sim 1$ eV, $N_{o}\sim 10^{9}$ cm$^{-3}$. In this case the\
Coulomb parameter for a negatively-charged dust particle in the presence of
the plasma sheath produced by hydrogen ions may result typically $\Gamma
_{di}\cong 0.91\times 10^{-6}\frac{Z_{q}\left( \frac{4\pi }{3}N_{o}\text{ cm}%
^{-3}\right) ^{1/3}}{\left( T\text{ eV}\right) }\cong 4\pi \times \left(
15\div 150\right) ,$ while the dimensionless radius of the ion plasma sheath
can be estimated $x_{oi}\cong 0.83\times 10^{-3}\frac{\left( N_{o}\text{ cm}%
^{-3}\right) ^{1/6}}{\sqrt{T\text{ eV}}}\cong 0.03.$ However, in principle,
in a dusty plasma the plasma Debye length can also become significantly
smaller than in the corresponding "clean" plasma (i.e., before injection of
the dusty species), which may influence, i.e., increasing, the values of $%
\Gamma _{di}$ and $x_{oi}$. Further important aspects of DSP concern the
investigation of possible charge screening effects, namely the reduction to
the effective charge $c$ with respect to the asymptotic value characterizing
the weak-field approximation (\ref{Eq.1-bbb}) of the DH potential (\ref%
{Eq.1-bb}). These effects, produced by non-linear contributions in the
Poisson equation, while usually negligible for weakly coupled plasmas, are
known to be significant in strongly-coupled plasmas \cite{Tessarotto 1992},
such as dusty plasmas \cite{Bystrenko 1999,Tsytovich 2001}. Indeed the
investigation of the effective interactions characterizing high-Z grains in
plasmas has attracted interest in recent years especially for numerical
simulations (see, for example, \cite{Robbins-1988,Allayrov 1998,Bystrenko
2003}). However, analytic estimates of the effective charge characterizing
the DH potential in strongly-coupled plasmas are still not available.

Goal of this work is the analysis of DSP and the definition of a suitably
modified mathematical model to take into account the effect of local plasma
sheaths in quasi-neutral plasmas. In particular, in Sec. 2 the traditional
DH model is recalled, while the new model is presented in Sec.3. In the same
section the basic mathematical results are given which concern the
asymptotic properties of the solutions obtained in the limit of vanishing
sheath radii ($\rho _{os}\rightarrow 0^{+}$). \ We intend to prove that the
Poisson equation characterizing the DH model (here denoted Debye-Poisson
equation) can be considered as a limit equation obtained from a suitably
modified formulation of DSP. The latter is obtained by including the effect
of finite local plasma sheaths (i.e., requiring $x_{os}>0,$ for all
species). Basic feature of the present approach is the representation of the
Poisson equation in integral form. This permits to analyze the asymptotic
properties of the solutions of the modified problem in the limit $%
x_{os}\rightarrow 0^{+}.$ It is found, that the limit solution of the
modified DSP for $x_{os}\rightarrow 0^{+}$ is a distribution which vanishes
identically for all $\rho >0$ and is discontinuous in $\rho =0.$ It is found
(see THM.1 in Sec.3) that in this limit the solution of the modified DSP
results uniquely determined as a function of the normalized charge carried
by the test particle ($\beta $) and is represented by a well-defined
distribution. Precisely it follows that (see THM.1 in Sec.3)%
\begin{equation}
\lim_{x_{o}\rightarrow 0^{+}}\sinh \widehat{\Phi }_{x_{o}}(x)=\frac{\beta }{%
x^{2}}\delta (x),  \label{Eq.3-zz}
\end{equation}%
while one can prove\ that the limit function\ $\lim_{x_{o}\rightarrow 0^{+}}%
\widehat{\Phi }_{x_{o}}(x)\equiv \widehat{\Phi }(x)$ reads (see THM.3 in
Sec.4)%
\begin{equation}
\lim_{x_{o}\rightarrow 0^{+}}\widehat{\Phi }_{x_{o}}(x)=\lim_{x_{o}%
\rightarrow 0^{+}}\widehat{\Phi }_{x_{o}}(x_{o})\exp \left\{
y^{(int)}(x,x_{o})\right\} ,  \label{Eq.3-zx}
\end{equation}%
where $y^{(int)}(x,x_{o})$ is a suitable smooth real function to be defined
[see Sec.4, Eq.(\ref{asympt. sol. for y(x)})] so that there results $%
y(x_{o},x_{o})=0$ while for $x\neq x_{o}$ the following limit holds%
\begin{equation}
\lim_{x_{o}\rightarrow 0^{+}}y^{(int)}(x,x_{o})=-\infty .  \label{Eq.3-xxx}
\end{equation}%
Hence, obviously, the limit function $\widehat{\Phi }(x)$ is not a strong
solution of the DP equation. This is therefore a characteristic property of
the DH model. In particular, as a basic consequence, the effective charge of
the DH asymptotic solution $c$ vanishes identically in such a limit and
results independent of the charge of the test particle.

As a further development, we intend to investigate strongly-coupled plasmas
(Sec.4 and 5), for which there results for a test particle of species $k$
(for example to be identified with a dusty grain)
\begin{align}
\Gamma _{ks}& \sim \frac{1}{O(\delta )}\gg 1  \label{Eq.4-a} \\
x_{os}& \sim O(\delta ^{k}),  \label{Eq.5-a}
\end{align}%
\ (\emph{strong coupling ordering}) where $k=0,1$ and $\delta $ is an
infinitesimal. The asymptotic solutions of the modified DSP are here
determined explicitly for the external and internal asymptotic solutions.
The former, which are valid outside the Debye sphere (i.e., in the domain $%
x>1,$ or $x\gg 1$) coincides with the customary DH potential. The latter
instead, which occurs inside the Debye sphere near the boundary of the local
plasma sheath, describes the non-linear effects due to the screening
generated by plasma species having opposite charge with respect to that of
the test particle. In addition, in Sec.4 also the asymptotic solution of the
modified DSP close to the boundary of the local plasma sheath (i.e., for $%
\Delta x=x-x_{os}$ infinitesimal) is constructed. It is proven that the
asymptotic solution can be determined uniquely, together with its internal
boundary conditions (defined on the boundary of the plasma sheath, i.e., for
$x=x_{os})$. Finally, by comparing the internal and external asymptotic
solutions for a test particle in a strongly-coupled plasma, an "asymptotic"
upper bound is obtained (Sec.5) for the effective charge characterizing the
DH potential. As a result, it is found that in strongly-coupled plasmas the
effective charge of highly-charged test particles, which characterizes the
DH asymptotic solution, results strongly reduced with respect the value of
the isolated test particle (\emph{charge screening effect})$.$ The result
appears potentially relevant especially for strongly-coupled dusty plasmas,
since the charge screening effect can dramatically reduce the DH potential,
particularly for highly charged test particles.

\section{The Debye-H\"{u}ckel model and the DSP problem}

The traditional formulation of the DSP, based on the
Debye-H\"{u}ckel model \cite{Debye 1923} regards the test
particles as point-like and having a spherically-symmetric charge
distribution while ignoring the effect of local plasma sheath.
This implies, from the physical standpoint, to neglect the
discrete nature of the plasma. As a consequence the background
plasma is treated as a continuum medium formed potentially by
several particle species, each described by a Maxwellian kinetic
distribution, so that its total charge density results as $\rho
_{p}=\sum\limits_{s}Z_{s}eN_{os}\exp
\left\{ -\frac{Z_{s}e\Phi }{2T_{s}}\right\} ,$where the constraint $%
\sum\limits_{s}Z_{s}eN_{os}=0$ must be imposed for quasi-neutral
plasma. Here the summation is performed on all particle species of
the plasma, while $N_{os}$ and $T_{s}$ (species density
coefficient and temperature) are both assumed constant.
Furthermore, the charge density of the test particles is
manifestly $\rho _{n}=\sum\limits_{i}\frac{q_{i}\delta (\left\vert \mathbf{%
r-r}_{i}(t)\right\vert )}{4\pi \left\vert \mathbf{r-r}_{i}(t)\right\vert ^{2}%
},$\ where $q_{i}\equiv Z_{q_{i}}e$ and $\mathbf{r}_{i}(t)$ are the electric
charge and position vector of the $i-$th test particle and the summation is
carried out on the collection of all test particles. In the remainder we
shall assume, without loss of generality, for all test charges $q_{i}>0.$
Hence, in terms of dimensionless electrostatic potential $\widehat{\Phi }=%
\frac{\left\vert e\right\vert \Phi }{T},$ one obtains, for example, $\rho
_{p}=2eN_{o}\sinh \widehat{\Phi },$ for a quasi-neutral electron-proton
plasma. In a case of a single point charge, in terms of the Debye length $%
\lambda _{D}=\sqrt{\frac{T}{8\pi e^{2}N_{o}}}$ and the dimensionless
distance $x=\rho /\lambda _{D}$ (where $\rho =\left\vert \mathbf{r-r}%
(t)\right\vert \in $), \ $\widehat{\Phi }(x)$ must satisfy the equation:\ \
\begin{equation}
\nabla _{x}^{2}\widehat{\Phi }(x)=-\frac{\beta \delta (x)}{x^{2}}+\sinh
\widehat{\Phi }(x),  \label{DSP equation 2b}
\end{equation}%
\ (\emph{DP equation}), where $\nabla _{x}^{2}\equiv \frac{1}{x^{2}}\frac{d}{%
dx}\left( x^{2}\frac{d}{dx}\right) $ (the Laplacian expressed in
dimensionless variables). We require that $\widehat{\Phi }(x)$ obeys the
boundary conditions, i.e.,%
\begin{align}
\lim_{x\rightarrow \infty }\widehat{\Phi }(x)& =0,  \label{bc-1} \\
\lim_{x\rightarrow 0}x^{2}\frac{d}{dx}\widehat{\Phi }(x\mathbf{)}& \mathbf{=-%
}\beta .  \label{bc-2}
\end{align}%
Equation (\ref{DSP equation 2b}), together with (\ref{bc-1}), (\ref{bc-2}),
define the so-called \emph{Debye screening problem} (\emph{DSP}). We notice
that, in contrast to what stated earlier \cite{Lo Surdo 1994}, both boundary
conditions (\ref{bc-1}), (\ref{bc-2}) can be satisfied by $\widehat{\Phi }%
(x).$ In fact, it is immediate to prove that in the domain $x>0,$ and
imposing (\ref{bc-1}), (\ref{bc-2}), the DP equation can be cast in the
integral form%
\begin{equation}
\widehat{\Phi }(x)\mathbf{=}\frac{\beta }{x}-\left[ \frac{1}{x}%
\int_{0}^{x}dx^{\prime }x^{\prime 2}+\int_{x}^{\infty }dx^{\prime
}x^{\prime }\right] \sinh \widehat{\Phi }(x^{\prime }).
\label{integral form of DSP equation}
\end{equation}%
A strong solution of (\ref{DSP equation 2b}),(\ref{bc-1}), (\ref{bc-2}) [or,
equivalent, of Eq.(\ref{integral form of DSP equation})] is a solution
belonging to the functional class $\widehat{C}^{(2)}(%
\mathbb{R}
^{+})\equiv \left\{ \widehat{\Phi }(x):C^{(0)}\left( \left[ 0,\infty \right[
\right) ,C^{(2)}\left( \left] 0,\infty \right[ \right) \right\} $. In the
traditional approach \cite{Debye 1923} the DSP problem is solved requiring
the validity for all $x\mathbf{\in }^{+}$ of the weak electric field ordering%
\emph{\ }(\ref{eq-1-a}) and assuming that $\widehat{\Phi }(x\mathbf{)=}%
\widehat{\Phi }_{o}(x)+O(\varepsilon )$ (\emph{weak field approximation}), $%
\widehat{\Phi }_{o}(x)$ being the solution of the linearized DP equation :

\begin{equation}
\nabla _{x}^{2}\widehat{\Phi }_{o}(x)=-\frac{\beta }{x^{2}}\delta (x)+%
\widehat{\Phi }_{o}(x),  \label{DSP equation 3}
\end{equation}%
with $\widehat{\Phi }_{o}(x),$ by assumption, belonging to the same
functional class of $\widehat{\Phi }(x)$. This equation admits the exact
solution $\widehat{\Phi }_{o}(x)=\frac{c}{x}e^{-x},$ i.e., Eq.(\ref{Eq.1-bb}%
), defined in terms of the DH effective charge $c=\beta .$ Finally, we
stress that, assuming the validity of the weak electric field ordering, due
to the linearity of Eq.(\ref{DSP equation 3}) a similar result is implied
also for a collection of point charge sources. In this case, in fact, the
corresponding solution of (\ref{DSP equation 3}) is just a sum of DH
potentials, one associated with each source.

\section{Consistency of the mathematical model - the modified DSP}

In this section we define the \emph{modified Debye screening problem}, based
on the introduction of a simple model of local plasma sheath. In the sequel
we shall consider for simplicity of notation the case of a two
species-plasma, formed by electrons and Hydrogen ions, having an unique
plasma sheath. However, the generalization to multispecies plasma is
straightforward. Thus, we shall assume that the test particle is represented
by a spherically symmetric charge of radius $\rho _{p}.$ For a particle in
which $\rho _{p}<\rho _{o}$ the plasma sheath is represented by the
spherical shell centered at the position (center) of the test particle for
which $\rho _{p}\leq \rho <\rho _{o}$, in which the plasma charge density
(except for the presence of the test particle) results negligible. In the
sequel we can also let in particular $\rho _{p}=0$ (point-like test
particle) or $\rho _{p}=\rho _{o}$ (finite-size test particle).

The customary DH model is obviously recovered letting $\rho _{p}=0$ and
taking the limit $\rho _{o}\rightarrow 0$ (or in dimensionless variables,
requiring $x_{p}\equiv \rho _{p}/\lambda _{D}=0$ and $x_{o}\equiv \rho
_{o}/\lambda _{D}$\ $\rightarrow 0$). Denoting $\widehat{\Phi }_{x_{o}}(x)$
the solution of the Poisson equation, here we intend to determine its
asymptotic properties in the limit $x_{o}\rightarrow 0^{+},$ while\ also
letting $x_{p}=0$ (see THM.1). As a consequence and in agreement with \cite%
{Lampert 1980,Garrett 1988}, in such a case it follows that the limit
function $\lim_{x_{o}\rightarrow 0^{+}}\widehat{\Phi }_{x_{o}}(x)\equiv
\widehat{\Phi }(x)$ vanishes identically for $x>0$
\begin{equation}
\left. \lim_{x_{o}\rightarrow 0^{+}}\widehat{\Phi }_{x_{o}}(x)=0.\right.
\label{Eq.3-2}
\end{equation}%
In addition, in the same set we intend to prove the identity
\begin{equation}
\left. \beta -\lim_{x_{o}\rightarrow
0^{+}}\int_{x_{o}}^{x}dx^{\prime
}x^{\prime 2}\sinh \widehat{\Phi }_{x_{o}}(x^{\prime })\widehat{\Theta }%
(x^{\prime }-x_{o})=0,\right.   \label{Eq.3-3}
\end{equation}%
which implies necessarily Eq.(\ref{Eq.3-zz}), $\widehat{\Theta }(x-x_{o})$
being the weak Heaviside function
\begin{equation}
\widehat{\Theta }(x-x_{o})=\left\{
\begin{array}{ccc}
0 &  & x<x_{o} \\
1 &  & x\geq x_{o.}%
\end{array}%
\right.
\end{equation}%
In detail the relevant equations valid in each subdomain for the normalized
electrostatic potential $\widehat{\Phi }_{x_{o}}(x\mathbf{)}$ are as
follows. In the internal domain $0\leq x<x_{p}$ the electrostatic potential
is assumed constant%
\begin{equation}
\widehat{\Phi }_{x_{o}}(x\mathbf{)=}\widehat{\Phi }_{x_{o}}(x_{p}\mathbf{).}
\end{equation}%
In the plasma sheath $x_{p}\leq x<x_{p},$ $\widehat{\Phi }_{x_{o}}(x)$
satisfies the customary Poisson equation in the presence of the charge
density produced by a finite-size spherically-symmetric charge
\begin{equation}
\nabla _{x}^{2}\widehat{\Phi }_{x_{o}}=-\frac{\beta }{x^{2}}\delta (x-x_{p}).
\label{DSP-modified -b}
\end{equation}%
Finally, in the external domain $x>x_{o}$ there holds the Poisson equation
in the presence of the plasma charge density:
\begin{equation}
\nabla _{x}^{2}\widehat{\Phi }_{x_{o}}=\widehat{\Theta }(x-x_{o})\sinh
\widehat{\Phi }_{x_{o}}.  \label{DSP modified}
\end{equation}%
The boundary conditions analogous to (\ref{bc-1}),(\ref{bc-2}), imposed
respectively at infinity and at the boundary of the plasma sheath, are
specified as follows
\begin{align}
\lim_{x\rightarrow \infty }\widehat{\Phi }_{x_{o}}(x\mathbf{)}& =0,
\label{BC-1 b} \\
\left. x^{2}\frac{d}{dx}\widehat{\Phi }_{x_{o}}(x\mathbf{)}\right\vert
_{x=x_{o}}& \mathbf{=-}\beta .  \label{BC-2 b}
\end{align}%
We notice that, if $x_{p}<x_{o}$ (for example, $x_{p}=0$)$,$ $\widehat{\Phi }%
_{x_{o}}(x\mathbf{)}$ results by assumption at least of class $C^{(1)}(%
\mathbb{R}
_{\{x_{p}\}}),$ where $%
\mathbb{R}
_{\left\{ x_{p}\right\} }\equiv \left] x_{p},\infty \right[ $ . Here $%
x_{o},\beta $ are both assumed constant and strictly positive real numbers.
The problem defined by (\ref{DSP modified}),(\ref{DSP-modified -b}),
together with the boundary conditions (\ref{BC-1 b}),(\ref{BC-2 b}), will be
here denoted as \emph{modified DSP. }\ From the physical standpoint Eqs.(\ref%
{DSP modified}),(\ref{DSP-modified -b}) may be viewed as the Poisson
equation for a spherical ideally conducting charge, or for a point particle
in the presence of a plasma sheath, of radius $r_{o}$ (i.e., $%
x_{o}=r_{o}/\lambda _{D}$ in non-dimensional variables) which is in
electrostatic equilibrium and is immersed in a spatially uniform
quasi-neutral and Maxwellian plasma. As for the previous DP equation, it
follows that, for solutions satisfying the boundary conditions (\ref{BC-1 b}%
),(\ref{BC-2 b}), in the domain $x\in
\mathbb{R}
_{\left\{ x_{o}\right\} }$ Eq.(\ref{DSP modified}) can be cast in the
integral form%
\begin{align}
\widehat{\Phi }_{x_{o}}(x)& \mathbf{=}\frac{\beta \widehat{\Theta }(x-x_{o})%
}{x}-  \label{integral form - DSP modified} \\
& -\left[ \frac{1}{x}\int_{x_{o}}^{x}dx^{\prime }x^{\prime
2}+\int_{x}^{\infty }dx^{\prime }x^{\prime }\right] \sinh \widehat{\Phi }%
_{x_{o}}(x^{\prime })\widehat{\Theta }(x^{\prime }-x_{o}).
\end{align}%
In particular, thanks to continuity at $x=x_{o}$ of $\widehat{\Phi }%
_{x_{o}}(x),$ one obtains the constraint
\begin{equation}
\widehat{\Phi }_{x_{o}}(x_{o})=\Gamma -\int_{x_{o}}^{\infty
}dx^{\prime
}x^{\prime }\sinh \widehat{\Phi }_{x_{o}}(x^{\prime })\widehat{\Theta }%
(x^{\prime }-x_{o}),  \label{constraint for Fi}
\end{equation}%
with $\Gamma \equiv \frac{\beta }{x_{o}}$ denoting the Coulomb coupling
parameter. It is immediate to establish, the existence and uniqueness of $%
\widehat{\Phi }_{x_{o}}(x)$ in the functional \ class $\widehat{C}^{(\infty
)}(%
\mathbb{R}
_{\left\{ x_{o}\right\} }),$ together with its continuous dependence on
initial data, in particular the continuity with respect to the parameter $%
x_{o}\in _{\left\{ 0\right\} }.$ Moreover, assuming that the weak-fields
approximation (\ref{eq-1-a}) applies (this condition is manifestly fulfilled
identically in the weak-coupling ordering,$\ 0<\Gamma \sim O(\varepsilon
)\ll 1),$ and is satisfied at least for $x\gg 1$ suitably large$,$ it is
immediate to prove that in this subset an asymptotic\ solution of the
modified DSP is provided by the \emph{external asymptotic solution}
\begin{equation}
\widehat{\Phi }_{x_{o}}(x)\cong \widehat{\Phi }_{x_{o}}^{(ext)}(x)\equiv
\frac{c}{x}e^{-x+x_{o}}.  \label{external asymptotic solutionl}
\end{equation}%
Here denoted as of the modified DSP and $c=c(x_{o},\Gamma )$ is the \emph{%
effective dimensionless charge.} Hence, $\widehat{\Phi }_{x_{o}}^{(ext)}(x)$
reduces formally to the previous DH potential (\ref{Eq.1-bb}) when\ $x\gg
x_{o}.$ In the weak-coupling ordering it follows $c(x_{o},\Gamma )=\frac{%
\beta }{1+x_{o}},$ while for strongly-coupled plasmas a lower value is
expected. Furthermore, it is obvious that the limit function $%
\lim_{x_{o}\rightarrow 0^{+}}\widehat{\Phi }_{x_{o}}(x)$ coincides with the
solution of DSP, i.e.,%
\begin{equation}
\lim_{x_{o}\rightarrow 0^{+}}\widehat{\Phi }_{x_{o}}(x)=\Phi (x).
\label{limit function and DSP}
\end{equation}

Let us now investigate the asymptotic properties of $\ $the exact solution $%
\widehat{\Phi}_{x_{o}}(x)$ in the limit $x_{o}\rightarrow0^{+}$.

\subsection{THEOREM 1 - Asymptotic properties of $\widehat{\Phi}_{x_{o}}(x)$}

\emph{For any strong solution of the modified DSP, }$\widehat{\Phi }%
_{x_{o}}(x)$ \emph{obtained letting} $x_{p}=0,$\emph{\ the limit function }$%
\lim_{x_{o}\rightarrow 0^{+}}\widehat{\Phi }_{x_{o}}(x)$\emph{\ has the
following properties:}

\emph{1) }%
\begin{equation}
\lim_{x_{o}\rightarrow0^{+}}\widehat{\Phi}_{x_{o}}(x_{o})=+\infty;
\label{limit-1}
\end{equation}

\emph{2) for any }$x>0$\emph{\ Eq.(\ref{Eq.3-2}) holds identically.}

\emph{3) for any }$x>0$ \emph{the integral limit (\ref{Eq.3-3}) is satisfied
by }$\widehat{\Phi}_{x_{o}}(x).$\emph{This implies that the limit function }$%
\widehat{\Phi}(x)=\lim_{x_{o}\rightarrow0^{+}}\widehat{\Phi}_{x_{o}}(x)$
\emph{results such that Eq.(\ref{Eq.3-zz}) is satisfied identically.
Moreover, for any }$x>0,x\in_{\left\{ 0\right\} }$%
\begin{equation}
\widehat{\Phi}(x)=\lim_{x_{o}\rightarrow0^{+}}\widehat{\Phi}_{x_{o}}(x)=0.
\label{limit-3c}
\end{equation}

\emph{4) the following limit is satisfied by the boundary value }$\widehat{%
\Phi}_{x_{o}}(x_{o})$%
\begin{equation}
\lim_{x_{o}\rightarrow0^{+}}x_{o}\widehat{\Phi}_{x_{o}}(x_{o})=0.
\label{limit-4a}
\end{equation}

\emph{5) the limit value of the effective dimensionless charge }$%
c(x_{o},\Gamma)$\emph{\ for }$x_{o}\rightarrow0^{+}$\emph{, obtained keeping
}$\Gamma$ \emph{finite, is}%
\begin{equation}
\lim_{x_{o}\rightarrow0^{+}}c(x_{o},\Gamma)=0.  \label{limit-5}
\end{equation}

PROOF

1) In fact, as a consequence of the integral equation (\ref{constraint for
Fi}) and the continuous dependence of $\widehat{\Phi }_{x_{o}}(x)$ on the
initial data, it follows%
\begin{equation}
\lim_{x_{o}\rightarrow 0^{+}}x_{o}\widehat{\Phi
}_{x_{o}}(x_{o})=\beta -\lim_{x_{o}\rightarrow
0^{+}}x_{o}\int_{x_{o}}^{\infty }dx^{\prime
}x^{\prime }\sinh \widehat{\Phi }_{x_{o}}(x^{\prime })\widehat{\Theta }%
(x^{\prime }-x_{o}),  \label{limit-4b}
\end{equation}%
which implies
\begin{align}
& \left. \lim_{x_{o}\rightarrow 0^{+}}\widehat{\Phi }_{x_{o}}(x_{o})=\infty
,\right. \\
& \left. \lim_{x_{o}\rightarrow 0^{+}}\int_{x_{o}}^{\infty
}dx^{\prime }x^{\prime }\sinh \widehat{\Phi
}_{x_{o}}(x)\widehat{\Theta }(x^{\prime }-x_{o})=\infty ,\right.
\end{align}%
i.e., the limit function $\lim_{x_{o}\rightarrow 0^{+}}\widehat{\Phi }%
_{x_{o}}(x)$ diverges in $x=x_{o}.$ Therefore, due to the continuity of \ $%
\widehat{\Phi }_{x_{o}}(x)$ with respect to $x\in \left[ 0,\infty \right[ $
it follows that infinitesimally close to $x,x_{o}=0,$ and when $x,x_{o}$ are
infinitesimal of the same order, $\widehat{\Phi }_{x_{o}}(x)$ must diverge
logarithmically as
\begin{equation}
\widehat{\Phi }_{x_{o}}(x)\sim \ln \left\{ \frac{1}{x^{3}}\right\} .
\label{divergency}
\end{equation}

2) Let us now consider the implications of the integral equation (\ref%
{integral form - DSP modified}) for the limit function \ $\lim
_{x_{o}\rightarrow0^{+}}\widehat{\Phi}_{x_{o}}(x)$ for arbitrary $%
x\in_{\left\{ x_{o}\right\} }.$ There follows%
\begin{align}
\lim_{x_{o}\rightarrow0^{+}}\widehat{\Phi}_{x_{o}}(x) & \mathbf{=}\frac{\beta%
}{x}-\frac{1}{x}\lim_{x_{o}\rightarrow0^{+}}\int
_{x_{o}}^{x}dx^{\prime}x^{\prime2}\sinh\widehat{\Phi}_{x_{o}}(x^{\prime})%
\widehat{\Theta }(x^{\prime}-x_{o})- \\
& -\int _{x}^{\infty}dx^{\prime}x^{\prime}\lim_{x_{o}\rightarrow0^{+}}\sinh%
\widehat{\Phi}_{x_{o}}(x^{\prime})\widehat{\Theta}(x^{\prime}-x_{o}),  \notag
\end{align}
where, due to the asymptotic estimate (\ref{divergency}), the second term on
the r.h.s. necessarily diverges
\begin{equation}
\lim_{x_{o}\rightarrow0^{+}}\frac{1}{x}\int
_{x_{o}}^{x}dx^{\prime}x^{\prime2}\sinh\widehat{\Phi}_{x_{o}}(x^{\prime})%
\widehat{\Theta }(x^{\prime}-x_{o})=\infty
\end{equation}
unless there results for any $x\neq x_{o},$ $x\in_{\left\{ x_{o}\right\} }$
\begin{equation}
\lim_{x_{o}\rightarrow0^{+}}\widehat{\Phi}_{x_{o}}(x)=0.  \label{limit-4 A}
\end{equation}

3) As a consequence of Eq.(\ref{limit-4 A}), from the integral equation (\ref%
{integral form - DSP modified}) it follows necessarily that for all $x>0$:%
\begin{equation}
\lim_{x_{o}\rightarrow0^{+}}\frac{\beta\widehat{\Theta}(x-x_{o})}{x}%
=\lim_{x_{o}\rightarrow0^{+}}\frac{1}{x}\int
_{x_{o}}^{x}dx^{\prime}x^{\prime2}\sinh\widehat{\Phi}_{x_{o}}(x^{\prime})%
\widehat{\Theta }(x^{\prime}-x_{o}),
\end{equation}
which proves the limit (\ref{Eq.3-3}). As a consequence it must result
necessarily that the limit $\lim_{x_{o}\rightarrow0^{+}}\sinh\widehat{\Phi }%
_{x_{o}}(x)$ is a Dirac delta, in accordance with Eqs. (\ref{Eq.3-2}) and (%
\ref{Eq.3-zz}).

4) The limit (\ref{limit-4a}) follows immediately from the boundary
condition (\ref{constraint for Fi}).

5) Finally, Eq.(\ref{limit-4 A}) implies manifestly the limit (\ref{limit-5}%
).

As an immediate consequence of THM.1 it follows that the DP equation of the
DH model does not admit strong solutions.

\subsection{THEOREM 2 - Non-existence of classical solutions of DSP}

\emph{In the functional class }$\widehat{C}^{(2)}(_{\left\{ 0\right\} })$%
\emph{\ the DSP problem defined by Eqs.(\ref{DSP equation 2b}),(\ref{bc-1})
and (\ref{bc-2}) has no strong solution.}

PROOF

In fact, first, we notice that the limit function%
\begin{equation}
\lim_{x_{o}\rightarrow0^{+}}\widehat{\Phi}_{x_{o}}(x)\equiv\widehat{\Phi}(x),
\end{equation}
is manifestly a solution of the DP equation which satisfies the required
boundary conditions (\ref{bc-1}), (\ref{bc-2}). Other hand, due to THM.1,
this solution is discontinuous in $x=x_{o}$ and results a distribution.
Hence it is not a strong (classical) solution of the modified DSP problem.

The basic implication of THM.1 and 2 is that the DP equation (\ref{DSP
equation 2b}), provided by the DH model, must be regarded as physically
unacceptable, since it does not admit strong solutions. In this regard it
should be noted that, as a basic principle, physically acceptable of
solutions of ordinary (or partial) differential equations characterizing the
classical theory of fields must be suitably smooth strong solutions. The
modified Debye screening problem here defined, instead, exhibits smooth
strong solutions and therefore appears, from this viewpoint, consistent.

\section{Internal asymptotic solution of the modified DSP}

In this section we intend to construct explicitly the internal asymptotic
solution of the modified DSP problem, defined by (\ref{DSP modified}),(\ref%
{DSP-modified -b}), (\ref{BC-1 b}),(\ref{BC-2 b}), in the case of
strongly-coupled plasmas, for which the asymptotic ordering defined by the
inequalities (\ref{Eq.4-a}),(\ref{Eq.5-a}) applies. In particular, we shall
consider a suitable neighborhood of the boundary of the plasma sheath ($%
x_{o} $), characterized by an infinitesimal amplitude $\Delta x=x-x_{o}$
defined so that
\begin{equation}
0\leq \Delta x\ll \frac{x_{o}w_{o}}{\Gamma }\sqrt{\ln w_{o}},  \label{Eq.1-4}
\end{equation}%
where by hypothesis $w_{o}\equiv \ln \left( \frac{2\Gamma ^{2}}{x_{o}^{2}}%
\right) \gg 1.$ Under such assumptions, we intend to construct approximate
solutions of the modified DSP which are accurate to leading order in the
infinitesimal dimensionless parameter $\delta $ and with respect to $\Delta
x.$These will be denoted as \emph{internal asymptotic solution }of the
modified DSP. We intend to prove, as a consequence, that in the neighborhood
of $x_{o}$ defined by (\ref{Eq.1-4}) the internal asymptotic solution can be
uniquely determined, to leading order in the relevant asymptotic parameter.
In fact, the following theorem holds.

\subsection{THEOREM 3 - Internal asymptotic solution for strongly-coupled
plasma}

\emph{In validity of the asymptotic ordering of strongly-coupled plasma [see
inequalities (\ref{Eq.4-a}),(\ref{Eq.5-a})], requiring }$x_{p}=0$\emph{\ (or
at least }$x_{p}\lesssim x_{o}$\emph{)} \emph{and in the neighborhood of }$%
x_{o}$\emph{\ defined by \ the inequalities (\ref{Eq.1-4}), the asymptotic
solution of the modified DSP [(\ref{DSP modified}),(\ref{DSP-modified -b}), (%
\ref{BC-1 b}),(\ref{BC-2 b})] reads }%
\begin{equation}
\widehat{\Phi }_{x_{o}}(x)\cong \widehat{\Phi }_{x_{o}}^{(int)}(x)=\widehat{%
\Phi }_{o}\exp \left\{ y^{(int)}(x,x_{o})\right\}  \label{Eq.1-4a}
\end{equation}%
\emph{where }$\widehat{\Phi }_{o}\equiv \widehat{\Phi }_{x_{o}}(x_{o})$\emph{%
\ and}%
\begin{equation}
y^{(int)}(x,x_{o})\cong y_{o}(x^{\prime },x_{o})\equiv -\alpha \left(
x-x_{o}\right)  \label{Eq.1-4aa}
\end{equation}%
\emph{(internal asymptotic solution), where }$\alpha =\frac{\Gamma }{x_{o}%
\widehat{\Phi }_{o}}$\emph{\ and the integration constant }$\widehat{\Phi }%
_{o}$\emph{\ results}
\begin{equation}
\widehat{\Phi }_{o}\cong \ln \left( \frac{2\Gamma ^{2}}{x_{o}^{2}}\right) .
\label{Eq.1-4-b}
\end{equation}

PROOF

In order to prove these results let us represent $\widehat{\Phi }_{x_{o}}(x)$
in the form%
\begin{equation}
\widehat{\Phi }_{x_{o}}(x)=\widehat{\Phi }_{o}\exp \left\{
y(x,x_{o})\right\} ,
\end{equation}

It follows that\ in the domain $x>x_{o},$ $y(x,x_{o})$ satisfies the
ordinary differential equation%
\begin{eqnarray}
&&\left. y"+\left( y^{\prime }\right) ^{2}+\frac{2}{x}y^{\prime }=\frac{1}{2%
\widehat{\Phi }_{o}}\exp \left\{ \widehat{\Phi }_{o}\exp \left\{ y\right\}
-y(x,x_{o})\right\} -\right.   \label{eq.DSP for w(x)} \\
&&\left. -\frac{1}{2\widehat{\Phi }_{o}}\exp \left\{ -\widehat{\Phi }%
_{o}\exp \left\{ y\right\} -y(x,x_{o})\right\} \equiv S(y(x,x_{o})),\right.
\end{eqnarray}%
where primes denote differentiation with respect to $x.$ Consistent with (%
\ref{BC-1 b}),(\ref{BC-2 b}), there follow the boundary conditions
\begin{equation}
y(x_{o},x_{o})=0,  \label{B-C-0}
\end{equation}%
\begin{equation}
\lim_{x_{o}\rightarrow 0^{+}}y(x,x_{o})=-\infty ,  \label{B-C-1}
\end{equation}%
\begin{equation}
\lim_{x\rightarrow \infty }y(x,x_{o})=-\infty ,  \label{BC-2}
\end{equation}%
\begin{equation}
\left. y^{\prime }(x,x_{o})\right\vert _{x=x_{o}}=-\frac{\Gamma }{x_{o}%
\widehat{\Phi }_{o}}.  \label{BC-3}
\end{equation}%
As a consequence, it is immediate to prove that $y(x,x_{o})$ for arbitrary $%
x,x_{o}>0$ ($x\geq x_{o})$ with admits the exact integral representation%
\begin{eqnarray}
&&\left. y(x,x_{o})=y_{o}(x,x_{o})+F(x,x_{o}),\right.
\label{int eq.for y(x)} \\
&&\left. y_{o}(x^{\prime },x_{o})=-\alpha (x-x_{o}).\right.   \label{yo(x)}
\end{eqnarray}%
Here $F(x,x_{o})$ is the solution of the integral equation
\begin{equation}
F(x,x_{o})=\int\limits_{x_{o}}^{x}dx^{\prime }\left( x-x^{\prime
}\right)
\left\{ S(y(x^{\prime },x_{o}))-y(x^{\prime },x_{o})^{2}-\frac{2}{x^{\prime }%
}y^{\prime }(x^{\prime },x_{o})\right\} ,  \label{Eq. for F(x,xo)}
\end{equation}%
while $S(y(x^{\prime },x_{o}))$ is defined by the r.h.s. of Eq.(\ref{eq.DSP
for w(x)}). Furthermore, a direct evaluation of the explicit integral in Eq.(%
\ref{Eq. for F(x,xo)}) permits to cast $F(x,x_{o})$ in the form%
\begin{eqnarray}
F(x,x_{o}) &=&F_{o}(x,x_{o})+F_{1}(x,x_{o}),
\label{int. eq. for F(x,xo) FINAL -1} \\
F_{1}(x,x_{o}) &=&\int\limits_{x_{o}}^{x}dx^{\prime }\left(
x-x^{\prime }\right) G_{1}(x^{\prime },x_{o}),
\end{eqnarray}%
where $F_{o}(x,x_{o})$ and $G_{1}(x^{\prime },x_{o})$ are respectively%
\begin{equation}
F_{o}(x,x_{o})=\int\limits_{x_{o}}^{x}dx^{\prime }\left(
x-x^{\prime
}\right) \left\{ S(y_{o}(x^{\prime },x_{o}))-y_{o}(x^{\prime },x_{o})^{2}-%
\frac{2}{x^{\prime }}y_{o}^{\prime }(x^{\prime },x_{o})\right\} ,
\label{int. eq. for F(x,xo) FINAL -0}
\end{equation}%
\begin{equation}
G_{1}(x^{\prime },x_{o})=S(y(x^{\prime },x_{o}))-S(y_{o}(x^{\prime
},x_{o}))-F_{1}(x^{\prime },x_{o})^{2}-2F_{1}(x^{\prime
},x_{o})y_{o}(x^{\prime },x_{o})-\frac{2}{x^{\prime }}F_{1}^{\prime
}(x^{\prime },x_{o}).  \label{int.eq. for F(x,xo) FINAL-2}
\end{equation}

Using Eqs.(\ref{int. eq. for F(x,xo) FINAL -1}), (\ref{int. eq. for F(x,xo)
FINAL -0}) and (\ref{int.eq. for F(x,xo) FINAL-2}) it is immediate to prove
that $y(x,x_{o})$ satisfies also the boundary conditions (\ref{B-C-0}),(\ref%
{BC-2}) and (\ref{BC-3}) and in particular that the limit (\ref{B-C-1}) is
actually satisfied. The proof follows by noting that, in the strong-coupling
ordering (\ref{Eq.4-a}),(\ref{Eq.5-a}) and for $0<\left( x-x_{o}\right) \sim
O(\delta ^{0}),$ the dominant contribution in $y(x,x_{o})$ is negative and
is provided by the second term on the r.h.s. of Eq.(\ref{int. eq. for
F(x,xo) FINAL -0}). This implies that an asymptotic approximation for $%
y(x,x_{o})$ $\ $is provided in this case by:%
\begin{equation}
y(x,x_{o})\cong y^{(int)}(x,x_{o})=-\alpha (x-x_{o})+F_{o}(x,x_{o}),
\label{asympt. sol. for y(x)}
\end{equation}%
and consequently also Eq.(\ref{Eq.1-4aa}). Manifestly $y^{(int)}(x,x_{o})$
satisfies identically all the boundary conditions (\ref{B-C-0}), (\ref{B-C-1}%
), (\ref{BC-2}) and (\ref{BC-3}). Finally, to determine the initial
condition $\widehat{\Phi }_{x_{o}}(x_{o})\equiv \widehat{\Phi }_{o},$ it is
sufficient to notice that in asymptotic orderings (\ref{Eq.4-a}),(\ref%
{Eq.5-a}) and (\ref{Eq.1-4}) the following identity must be satisfied,
correct to leading order in $\delta ,$
\begin{equation}
\beta \cong \frac{1}{2}\int_{x_{o}}^{x}dx^{\prime }x^{\prime
2}\exp \left\{ \widehat{\Phi }_{o}\exp \left\{ y(x^{\prime
},x_{o})\right\} \right\} . \label{Eq.10-4}
\end{equation}%
The integral can be estimated asymptotically to yield
\begin{equation}
\widehat{\Phi }_{x_{o}}(x_{o})\cong \ln \frac{2\Gamma }{x_{o}^{2}}\left[
\Gamma -\widehat{\Phi }_{x_{o}}(x_{o})\right] ,
\label{internal asymptotic solution}
\end{equation}%
which to leading order in $\delta $ implies Eq.(\ref{Eq.1-4-b}). To briefly
comment these results, it is important to remark, that, by assumption, the
internal asymptotic solution here obtained, (\ref{asympt. sol. for y(x)}),
is valid if the parameter $x_{o}$ results of order $O(1)$ or is
infinitesimal in $\delta $ [see the ordering (\ref{Eq.5-a})]$,$ while the
parameter $\Gamma $ satisfies (\ref{Eq.4-a})$.$ As a consequence, in the
limit $x_{o}\rightarrow 0^{+}$ Eq.(\ref{Eq.10-4}) results consistent with
Eq.(\ref{Eq.3-zx}), thus proving also the validity of Eq.(\ref{Eq.3-zz}).
Finally, we notice that the sub-domain (\ref{Eq.1-4}) in which the internal
asymptotic solution (\ref{asympt. sol. for y(x)}) is valid, allows $\widehat{%
\Phi }_{x_{o}}(x)\ll 1$. This is also consistent with the weak-field
approximation (\ref{eq-1-a}).

\section{Conclusions: asymptotic upper bound for the effective charge}

As an important application, the previous analytic results can be used to
obtain an asymptotic estimate for the effective dimensionless charge $%
c(x_{o},\Gamma )$ carried by the DH potential in strongly-coupled plasmas.
We intend to show that test particles which have a high charge undergo a
strong charge screening effect, close to the local plasma sheath, produced
by non-linear effects in the Poisson equation. As a consequence, outside the
Debye sphere (i.e., for $x>1)$ the DH potential generated by highly charged
test particles in strongly-coupled plasmas results strongly reduced with
respect to the theoretical value observed in the corresponding
weakly-coupled systems.

The effective charge of the test particle can be estimated by comparing the
internal and external asymptotic solutions in a suitable neighborhood of the
boundary of the plasma sheath (i.e., near $x=x_{o}$), belonging to the
sub-domain defined by (\ref{Eq.1-4}). For this purpose we impose the
matching condition
\begin{equation}
\widehat{\Phi }_{x_{o}}^{(int)}(x_{c})\cong \widehat{\Phi }%
_{x_{o}}^{(ext)}(x_{C}),  \label{Eq.1-5}
\end{equation}%
where, by assumption, $x_{C}$ is defined so that there results $\widehat{%
\Phi }_{x_{o}}(x_{c})\ll 1$ and moreover
\begin{equation}
\frac{\Gamma }{x_{o}\widehat{\Phi }_{o}}\Delta x_{c}\sim O(\delta )\ll 1.
\label{Eq.2-5}
\end{equation}%
being $\ \Delta x_{c}\equiv x_{c}-x_{o}$ and where $\frac{\Gamma }{x_{o}%
\widehat{\Phi }_{o}}\gg 1,$ consistent with the assumption of
strongly-coupled plasma (\ref{Eq.4-a}),(\ref{Eq.5-a}). The
matching condition can be used to obtain an upper estimate for
$c(x_{o},\Gamma ).$ In
fact, in validity of the ordering Eq.(\ref{Eq.2-5}), $\widehat{\Phi }%
_{x_{o}}^{(int)}(x_{c})$ can be approximated by \ $\widehat{\Phi }%
_{x_{o}}^{(int)}(x_{c})\cong \widehat{\Phi }_{o}\exp \left\{ -\frac{\Gamma }{%
x_{o}\widehat{\Phi }_{o}}(x-x_{o})\right\} .$ Then, ignoring higher order
corrections in $\delta ,$ the following inequality manifestly holds:%
\begin{equation}
c(x_{o},\Gamma )\lesssim c^{(a)(}(x_{o},\Gamma )\equiv x_{o}\ln \frac{%
2\Gamma ^{2}}{x_{o}^{2}},  \label{asymptotic estimate for c}
\end{equation}%
which delivers therefore an \emph{asymptotic upper bound }for the effective
charge $c(x_{o},\Gamma )$. In particular, it follows that $%
c^{(a)(}(x_{o},\Gamma )$ depends logarithmically on the normalized charge of
the isolated test particle $\beta $. \emph{Thus, in general for
strongly-coupled plasmas the effective dimensionless charge appears much
smaller than in weakly-coupled plasmas}. The result can be significant for
the investigation of dusty plasmas, particularly to describe the charge
screening effect of highly charged dusty grains. Comparisons with numerical
simulations, based on the\ numerical solution of modified DSP [defined by
Eqs. (\ref{DSP modified}),(\ref{DSP-modified -b}) and the boundary
conditions (\ref{BC-1 b}),(\ref{BC-2 b})], indicate that the asymptotic
estimate $c^{(a)(}(x_{o},\Gamma )$ holds already for $\beta \gtrsim 1$ and $%
x_{o}\lesssim 0.3$. For numerical calculations, the effective charge $%
c(x_{o},\Gamma )$ has been determined numerically as the limit for $x\gg 1$
(i.e., $x\rightarrow \infty $) of the the function
\begin{equation}
f_{c}(x)=\frac{\widehat{\Phi }_{x_{o}}(x)xe^{x-x_{o}}}{\beta },
\end{equation}%
where $\widehat{\Phi }_{x_{o}}(x)$ denotes the numerical solution of the
modified DSP. As previously discovered \cite{Tessarotto 1992}, it is found
that for strongly-coupled plasmas (i.e., assuming $\Gamma >1$) $f_{c}(x)$
approaches rapidly the asymptotic value $c(x_{o},\Gamma ),$ even at a
"distance" $x$ smaller than the Debye Length (i.e., for $x<1$). For
reference in figs 1-3 the numerical estimates of $f_{c}(x),c(x_{o},\Gamma )$
and $c^{(a)(}(x_{o},\Gamma )$ have been reported for the cases $\beta
=1,5,10 $ and $x_{o}=0.05\div 0.3.$ Its is found that the upper bound $%
c^{(a)(}(x_{o},\Gamma )$ provided by the majorization (\ref{asymptotic
estimate for c}) can actually be used to obtain an approximate estimate of
the effective charge $c(x_{o},\Gamma )$.

Finally, it is worthwhile pointing out that the asymptotic estimate provided
by $c^{(a)(}(x_{o},\Gamma )$\ satisfies also the correct limit set by THM.1.
In fact the it results
\begin{equation}
\lim_{x_{o}\rightarrow 0^{+}}c(x_{o},\Gamma )=0
\end{equation}%
and this limit is satisfied even in the case in which $\Gamma \sim 1/\delta $
$\sim 1/x_{o}.$

\section*{Acknowledgments}

Work supported by PRIN Research Program \textquotedblleft \textit{Programma
Cofin 2004: Modelli della teoria cinetica matematica nello studio dei
sistemi complessi nelle scienze applicate}\textquotedblright ( MIUR Italian
Ministry), the ICTP/TRIL Program (ICTP, Trieste, Italy), and the Consortium
for Magnetofluid Dynamics, University of Trieste, Italy.

\pagebreak

\begin{center}
\textbf{Figure captions}
\end{center}

\noindent \textbf{Figure 1 -} Comparison between $f_{c}(x),c(x_{o},\Gamma )$
and $c^{(a)(}(x_{o},\Gamma ).$ \ The data are normalized with respect to $%
\beta ,$ the normalized charge of the isolated test particle. The figure
concerns the case with $\beta =1$ and $x_{o}=0.05,$ yielding $\Gamma =20.$
The horizontal straight line represents the asymptotic estimate $%
c^{(a)(}(x_{o},\Gamma ),$ while the curve below it is the graph of $%
f_{c}(x). $ It follows $f_{c}(x_{o})\cong $ $0.564,$ while the asymptotic
value $c(x_{o},\Gamma )$ $\cong 0.493$ is reached approximately at $x\approx
0.3$, and the upper bound for the normalized effective charge is $%
c^{(a)(}(x_{o},\Gamma )\cong 0.589$. \newline

\noindent\textbf{Figure 2 -} Comparison between $f_{c}(x),c(x_{o},\Gamma )$
and $c^{(a)(}(x_{o},\Gamma )$ for $\beta =5$ and $x_{o}=0.2$ (with $\Gamma
=25$)$.$ In this case $f_{c}(x_{o})\cong $ $0.369,$ while the asymptotic
value $c(x_{o},\Gamma )$ $\cong 0.28$ is reached approximately at $x\approx
0.4$, and $c^{(a)(}(x_{o},\Gamma )\cong 0.38$. \newline

\noindent\textbf{Figure 3 -} Comparison between $f_{c}(x),c(x_{o},\Gamma )$
and $c^{(a)(}(x_{o},\Gamma )$ for $\beta =10$ and $x_{o}=0.3$ (with $\Gamma
\cong33$)$.$ In this case it is found $f_{c}(x_{o})\cong $ $0.273,$ while
the asymptotic value $c(x_{o},\Gamma )$ $\cong 0.188$ is reached
approximately at $x\approx 0.5$, and $c^{(a)(}(x_{o},\Gamma )\cong 0.303.$

\pagebreak
\begin{figure}[tbp]
\begin{center}
\includegraphics[width=.5\textwidth]{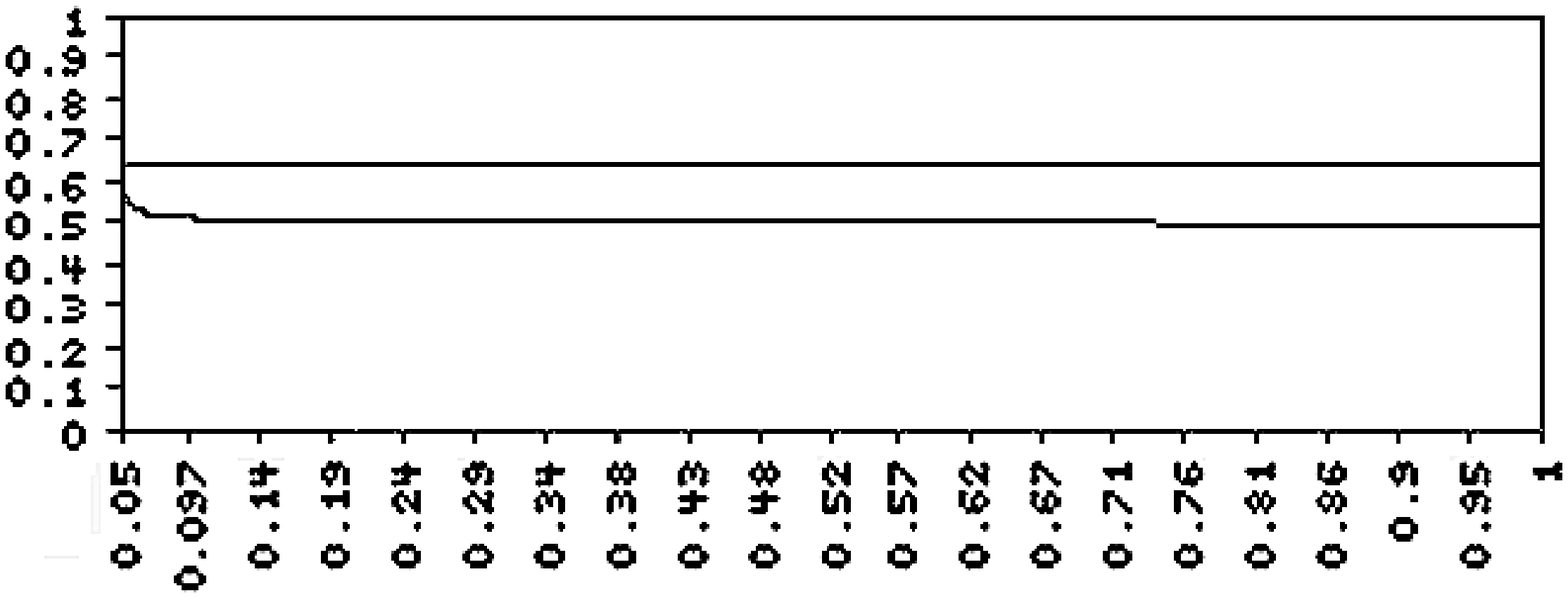}
\end{center}
\par
\vspace{-10pt}
\caption{}
\label{fig:1}
\end{figure}

\begin{figure}[tbp]
\begin{center}
\includegraphics[width=.5\textwidth]{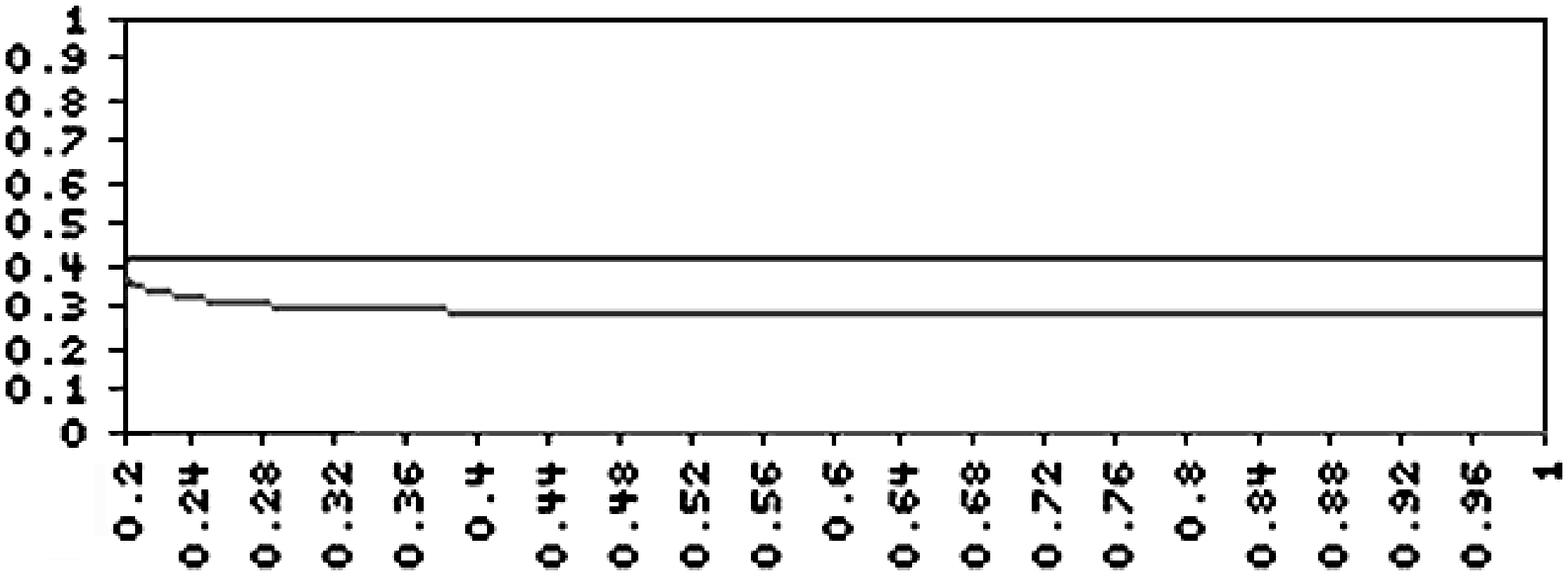}
\end{center}
\par
\vspace{-10pt}
\caption{}
\label{fig:2}
\end{figure}

\begin{figure}[tbp]
\begin{center}
\includegraphics[width=.5\textwidth]{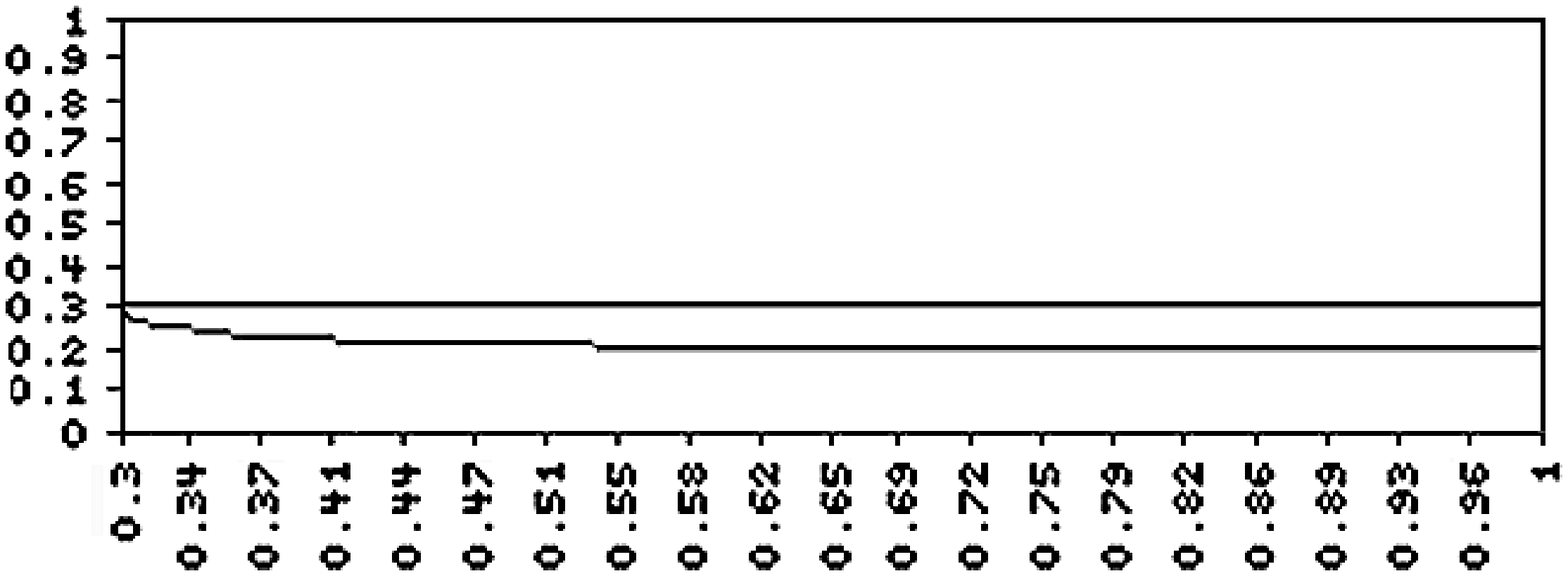}
\end{center}
\par
\vspace{-10pt}
\caption{}
\label{fig:3}
\end{figure}

\end{document}